\documentclass[aps,prb,twocolumn,superscriptaddress]{revtex4-2}
\usepackage{graphics,graphicx, array, afterpage}
\usepackage{qcircuit,amsmath}
\usepackage{grffile}
\usepackage[export]{adjustbox}
\usepackage{lineno}
\usepackage{xcolor}
\usepackage{longtable}
\usepackage{braket}
\usepackage{array}
\usepackage{multirow}
\usepackage{enumerate}
\usepackage{amsmath}
\usepackage{amssymb}
\usepackage{amsthm}
\usepackage{times}
\usepackage{multirow}

\definecolor{LinkColor}{RGB}{0,0,128}
\usepackage[colorlinks, citecolor=blue]{hyperref}
\hypersetup{linkcolor=LinkColor,urlcolor=LinkColor,citecolor=LinkColor,pdfstartview={FitH},urlcolor=LinkColor}

\usepackage[utf8]{inputenc}
\usepackage{color}
\usepackage{tabularx}
\usepackage{algorithm}
\usepackage{algpseudocode}
\usepackage{graphbox}
\usepackage{amsfonts}
\usepackage{dcolumn}
\usepackage{epstopdf}
\usepackage{lipsum}
\usepackage{amsmath,epsfig,amssymb,subfigure,bm,dsfont}
\graphicspath{{figures}}

\begin{document}
\preprint{APS/123-QED}

\title{Quantum feedback induced entanglement relaxation and dynamical phase transition in monitored free fermion chains with Wannier-Stark ladder}

\author{Xuyang Huang}
\thanks{These two authors contributed equally to this work.}
\affiliation{School of Physics and Optoelectronics, Xiangtan University, Xiangtan 411105, China}
\affiliation{Institute for Quantum Science and Technology, Shanghai University, Shanghai 200444, China}

\author{Han-Ze Li}
\thanks{These two authors contributed equally to this work.}
\affiliation{Institute for Quantum Science and Technology, Shanghai University, Shanghai 200444, China}

\author{Yu-Jun Zhao}
\affiliation{School of Physics and Optoelectronics, Xiangtan University, Xiangtan 411105, China}
\affiliation{Institute for Quantum Science and Technology, Shanghai University, Shanghai 200444, China}

\author{Shuo Liu}
\affiliation{Institute for Advanced Study, Tsinghua University, Beijing 100084, China}

\author{Jian-Xin Zhong}
\email{jxzhong@shu.edu.cn}
\affiliation{Institute for Quantum Science and Technology, Shanghai University, Shanghai 200444, China}
\affiliation{School of Physics and Optoelectronics, Xiangtan University, Xiangtan 411105, China}

\date{\today}

\begin{abstract}
In recent years, measurement induced entanglement transitions (MIETs) have attracted significant attention. However, the dynamical transition associated with the feedback induced skin effect, which exhibits a wealth of intriguing phenomena, has not been fully understood. In this work, we investigate a dynamical phase transition in a tilted free-fermion chain under measurement-feedback protocols, emphasizing the particle density and entanglement entropy dynamics. We reveal a feedback induced skin effect, enhanced by the Wannier-Stark ladder potential, that creates localization at one boundary and generates an effective pseudo edge under periodic conditions. The observables show a two-stage evolution: a rapid initial logarithmic growth followed by decay into an area-law steady state. Using a rescaling analysis, we pinpoint the critical behavior and offer an intuitive physical picture that links it to the feedback-driven suppression of quantum jump fluctuations. The resulting entanglement dynamics appear to be governed by a system-size-dependent delay, followed by a size-independent relaxation process. This behavior is consistent with the ballistic propagation of free fermions toward a domain-wall-like steady state and does not exhibit any signatures of nontrivial criticality. This work provides an effective supplement to the dynamical transition.  It provides valuable references for linking the dynamical understanding of the role feedback plays in MIETs.
\end{abstract}

\maketitle

\section{Introduction}

Depicting the evolution of quantum matter far from equilibrium with universal quantum dynamics might be notoriously challenging. However, considerable efforts to understand out-of-equilibrium dynamics in lattice models~\cite{Nandkishore2015, RevModPhys.91.021001, PhysRevLett.123.090603, 10.21468/SciPostPhysLectNotes.18, Moudgalya_2022, PhysRevA.108.062215, PhysRevLett.130.120403, liu2024quantummpembaeffectsmanybody,  PhysRevA.108.043301, PhysRevB.110.094310}, random unitary circuits~\cite{Nahum2017, 18PhysRevB.98.205136,Skinner2019, Jian2020, PhysRevX.7.031016,Potter2022, Fisher2023,PhysRevB.107.L201113, PhysRevLett.132.240402,PhysRevB.110.064323,PhysRevLett.133.140405},
black hole information theory~\cite{Patrick_Hayden_2007, Hartman2013, PhysRevD.106.086019}, and quantum field theory~\cite{Calabrese_2005, Calabrese_2009, PhysRevLett.112.011601} reveal that universal patterns may emerge in the entanglement structures of many-body systems. The core issues in condensed matter physics today, such as quantum thermalization~\cite{Nandkishore2015,doi:10.1126/science.aaf6725, Ueda2020}, quantum chaos~\cite{ PhysRevLett.98.044103, 10.21468/SciPostPhys.11.2.034, PhysRevResearch.6.033286}, holography~\cite{PhysRevLett.112.011601, Bhattacharyya2022}, and information scrambling~\cite{Landsman2019, PhysRevX.11.021010, PhysRevLett.129.050602}, are all characterized by distinct entanglement patterns.

In unitary entanglement relaxation dynamics (hereafter refer to entanglement relaxation), local perturbations spread entanglement within the Lieb-Robinson light cone~\cite{Lieb1972}, initially increasing entanglement entropy until it saturates. Growth patterns vary by system type: integrable systems exhibit slower growth~\cite{Vincenzo2018, PhysRevB.97.245135, Pasquale2020:, Piroli2020, PhysRevResearch.4.043212}, non-integrable systems thermalize and show faster growth~\cite{Adam2016, RevModPhys.91.021001}, and many-body localized systems display logarithmic growth due to localization effects~\cite{Bardarson2012, RevModPhys.91.021001, Schulz2019}. Interestingly, introducing weak measurements leads to a competition between unitary and nonunitary dynamics in the system, resulting in diverse entanglement relaxation patterns as well, thereby driving measurement-induced entanglement transitions (MIET)~\cite{Nahum2017, 18PhysRevB.98.205136,Skinner2019, Chan2019, Xiangyu2019, Szyniszewski2019, Bao2020,Ippoliti2021,Fisher2023}.

On the one hand, in the no-click limit~\cite{Yuto2020}, MIET corresponds to purely non-Hermitian dynamics, where the post-selection mechanism of measurement feedback effectively erases trajectories with quantum jumps. This phenomenon often directly reduces to an entanglement dynamics problem involving a Hatano-Nelson type~\cite{hatano1996,hatano1997,hatano1998} effective Hamiltonian, where entanglement relaxation is accompanied by a steady-state entanglement phase transition, eventually equilibrating to a non-Hermitian skin state~\cite{PhysRevX.13.021007,shen2024observationnonhermitianskineffect}. The relaxation of entanglement and MIET without quantum jumps have been extensively studied in free lattice models\cite{li2023disorderinducedentanglementphasetransitions, PhysRevB.109.024306, PhysRevB.110.094310, chen2024quantumentanglementnonhermiticityfreefermion, Youenn2023, PhysRevResearch.2.033069}, even including its interaction with localization potentials~\cite{li2023disorderinducedentanglementphasetransitions, PhysRevB.109.024306,zhou2024, PhysRevB.110.094310}. However, implementing post-selection experimentally is notably difficult. On the other hand, entanglement relaxation and MIET under quantum jumps~\cite{QJ} can also be described through stochastic dynamics in quantum many-body trajectories. In this context, if the effective Hamiltonian exhibits many-body skin pumping, quantum jump noise drives the entanglement steady state into a highly entangled configuration, rather than allowing it to settle into a low-entanglement non-Hermitian skin state. Consequently, the information scrambling from quantum jumps makes it challenging to observe the many-body skin effect in experimental settings.

However, substantial recent efforts have been made to understand entanglement relaxation with quantum jumps under generalized monitoring~\cite{sm3,wyp,PhysRevResearch.6.013244,Puente2024quantumstate}, especially within the context of the skin effect, in lattice models and quantum circuits~\cite{PhysRevLett.124.010603, fx, wyp, lzc, PhysRevB.110.144305}. There, feedback operations suppress the dynamical noise caused by quantum jumps, allowing the system's long-time steady-state to be dominated by the skin effect of the non-Hermitian effective Hamiltonian. Under the rescaled time $t/L$, where $L$ is the system size, the feedback-induced skin effect drives nonmonotonic bipartite entanglement growth, which has been identified as a novel dynamical transition~\cite{lzc}. However, several questions remain open following Ref.~\cite{lzc}: Is this truly a dynamical phase transition?  What is the underlying physical mechanism of this transition? Does this transition depend on the rescaling of  $t/L$?

More interestingly, the tilt system realized by introducing a linear potential into the Hamiltonian has emerged as a new playground for exploring exotic quantum states. Instead of undergoing unbounded acceleration, as one might expect from classical intuition, particles in a Wannier-Stark ladder can exhibit Bloch oscillations~\cite{Bloch1929,PhysRev.117.432,RevModPhys.63.91} or become localized in Wannier-Stark states~\cite{PhysRevLett.76.4508,PhysRevB.108.064206,zhao2024}. In many-body settings, tilts can lead to Stark many-body localization~\cite{Schulz2019,pnas,Morong2021} and Hilbert space fragmentation~\cite{titl1,Adler2024,hsf}, thereby hindering thermalization and giving rise to rich non-equilibrium phenomena. Notably, with the advent of increasingly clean quantum simulation platforms~\cite{Morong2021,Adler2024,hsf,pnas}, tilt systems come with a lower experimental overhead compared to other localization potentials. A natural question that arises is whether the aforementioned dynamical transition~\cite{lzc} could emerge in the presence of a Wannier-Stark ladder, potentially giving rise to even more intriguing nonequilibrium phenomena.


In this work, we investigate a dynamical phase transition in a tilted (Wannier-Stark laddered) free-fermion chain under monitoring-feedback protocols, with a focus on the dynamics of particle density profiles and entanglement entropy. We discover a feedback-induced skin pumping effect, where the presence of the Wannier-Stark ladder extends the evolution time required for the initial density profile to transition into the final domain-wall configuration. Under periodic boundary conditions, this potential generates an effective boundary or pseudo-edge, facilitating similar localization even in the absence of physical edges. Our analysis of entanglement entropy indicates a two-step evolution: a rapid initial increase following logarithmic scaling, subsequently transitioning into a gradual decay toward an area-law entangled steady state. Using multiple observables combined with data collapses, we accurately pinpoint the critical transition separating these dynamical regimes. Furthermore, we offer an intuitive picture connecting this behavior to feedback-driven suppression of quantum jump fluctuations. Importantly, the Wannier-Stark ladder delays system relaxation but does not alter the dynamical phase transition's critical behavior. Nevertheless, strong gradient potentials can independently lead the system directly into area-law entanglement, thus circumventing an explicit dynamical transition. This work effectively complement the unresolved issues regarding the mechanism and criticality of dynamical phase transitions highlighted in Ref.~\cite{lzc}. It provides valuable references for the possible connection between MIET and non-Hermitian dynamics.

This paper is organized as follows. Section~\ref{setup} introduces the model along with the monitoring-feedback protocol. 
In Section~\ref{nr},  we present a comprehensive analysis of the dynamical relaxation behavior in our system. We begin by examining the trajectory-averaged relaxations, focusing on the time evolution of the average density distribution. Next, we provide numerical evidence for the entanglement dynamical transitions under varying tilts, and discuss the evolution of other relevant observables during the relaxation process. Finally, using rescaling analysis,  we analyze the critical points of this dynamical phase transition, which further confirms the universal nature of the observed behavior.
Section~\ref{argument}  provides an explanation for this dynamical phase transition, summarizes the paper, and outlines future research directions. Additional numerical calculation data and discussions are included in the Appendix.

\begin{figure}[bt]
\hspace*{-0.5\textwidth}
\includegraphics[width=0.5\textwidth]{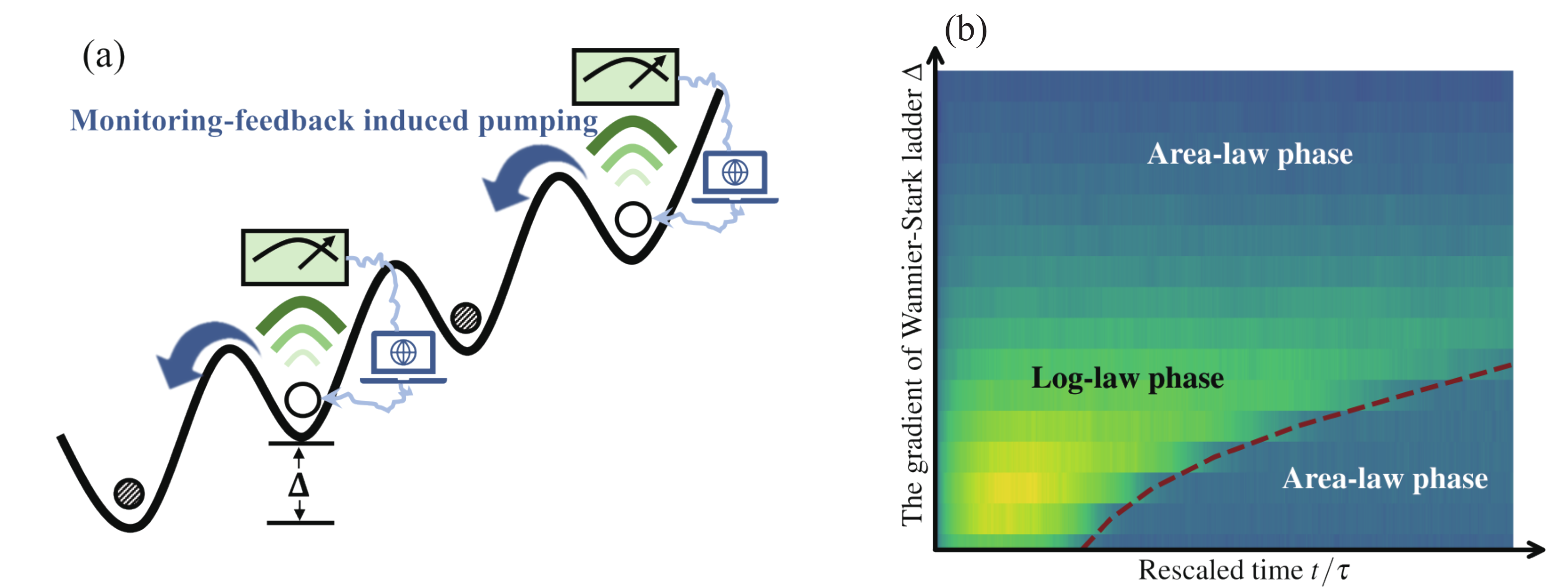} 
\caption{\label{smooth} (a) The influence of the continuous measurement with the feedback under the Wannier-Stark ladder. (b) OBCs, the dynamical phase diagram of the trajectory-averaged bipartite entanglement entropy in the plane ($t/\tau$, $\Delta$) with $L=192$. The figure is primarily segmented by log-law region and area-law region.}
\label{fig1}
\end{figure}

\section{Setups}\label{setup} 

\subsection{Model}

In this work, we consider a free fermion chain of length $L$ with Wannier-Stark ladder. The system is described by the following Hamiltonian, 
\begin{equation} \label{Eq:Hamiltonian}
    \hat{H} =  \sum_{l}{J(\hat{c}_l^\dagger \hat{c}_{l+1} + H.c.)} + \Delta\sum_{l}{ l \hat{c}_l^\dagger \hat{c}_l},
\end{equation}
where $\hat{c}_l^\dagger$ ($ \hat{c}_l$) is the fermion creation (annihilation) operator at site $l$,
$J$ is the strength of nearest-neighbor hopping which is set to $1$ hereafter for convenience, and $\Delta$ is the gradient of the Wannier-Stark ladder, i.e., the on-site potential increases linearly with the site index $l$ with the potential difference between the nearest-neighbor sites being $\Delta$.
Introducing the continuous measurements, the evolution of the open quantum system is described by the Lindblad master equation, which  is a formidable challenge for numerical simulation. Further, intrinsic property of  the trajectory state is not visible in the properties of the average state.

Thus, to simplify this process, we instead consider the SSE~\cite{QJ, SSE1, SSE2, SSE3, SSE4, OQS} corresponding to the trajectory evolution of a pure state $\left | \psi(t)  \right \rangle$  
\begin{equation} \label{Eq:SSE}
\begin{aligned}
    d\left | \psi(t)  \right \rangle
    &\ =-i \hat{H}_\text{eff} \left | \psi(t)  \right \rangle dt \\
    &\ +\sum_{l}^{L-1}\left(\frac{\hat{L}_l \left | \psi(t)  \right \rangle}
    {\sqrt{  \left \langle \hat{L}_l^\dagger \hat{L}_l \right \rangle }} -
   \left | \psi(t)  \right \rangle\right)  dW_l,
\end{aligned}
\end{equation} 
where $\hat{L}_l$ is a quantum jump operator induced by weak measurement, $  \langle  \hat{O}   \rangle   = \left\langle\psi(t)\right|\hat{O} \left | \psi(t)  \right \rangle $ is the expectation value of operator $\hat{O}$,
$dW_l$ is an independent and discrete Poisson random variable, which takes values from \{0, 1\} with mean value $ \langle \hat{L}_l^\dagger \hat{L}_l \rangle \gamma dt$.  $\hat{H}_\text{eff} $ is non-Hermitian effective Hamiltonian, 
$\hat{H}_\text{eff} = \hat{H} - i\gamma/{2} {\textstyle \sum_{l}^{L-1}{\hat{d}_l^\dagger \hat{d}_l}} $, where $\gamma$ is the monitoring rate and $\hat{d}^\dagger_l =  ( \hat{c}^\dagger_l - i \hat{c}^\dagger _{l+1}   )/\sqrt{2}$ is the creation operator of a right-moving quasimode.

We consider the continuous measurements with feedback, inspired by Refs.~\cite{wyp}. 
The quantum jump operator is written as 
\begin{equation} \label{Eq:qj}
    \hat{L}_l = e^{i\theta \hat{n}_{l+1}}\hat{d}_l^\dagger \hat{d}_l,
\end{equation}
where $e^{i\theta \hat{n}_{l+1}}$ is the feedback (see Appendix~\ref{edgefeedback} for boundary effect discussions), $\hat{n}_{l+1}$ is the particle number operator. If $\theta \ne 0$, the feedback will convert the right-moving quasimode  to
the left-moving one, and we set $\theta = \pi$ without loss of generality (for more discussion on theta, please refer to Appendix~\ref{theta}). 
Based on the previous text, we now can write $\hat{H}_\text{eff}$~[see FIG.~\ref{fig1} (a)] as
\begin{equation} 
\begin{aligned}
    \hat{H}_\text{eff} &\ = \sum_{l}^{L-1}{[(1+\frac{\gamma}{4})\hat{c}_l^\dagger \hat{c}_{l+1}
   +(1-\frac{\gamma}{4})\hat{c}_{l+1}^\dagger \hat{c}_{l}]}\\
   &\  -i \frac{\gamma}{4}\sum_{l}^{L-1}{(\hat{n}_l + \hat{n}_{l+1})}
    +\Delta \sum_{l}^{L}{ l \hat{n}_l}.
    \end{aligned}\label{Eq:Heff}
\end{equation}
In the subsequent sections, we set $\gamma=0.5$ unless otherwise specified, which corresponds to the case of the critical entanglement phase caused by the weak measurement~\cite{yaodongli}.

We choose the N\'e{e}l state as the initial state, i.e., $\left | \psi(t=0)  \right \rangle =\left | 0101...01  \right \rangle $. 
Since both $\hat{H}_\text{eff}$ and $\hat{L}_l$ are quadratic, 
we can utilize the Gaussian state to efficiently simulate each trajectory evolution of $\left | \psi(t)  \right \rangle$.
We note the observables are averaged 200-800 trajectories, represented by the variable with an overline (e.g., trajectory-averaged density distribution $\overline{ \langle \hat{n}_l \rangle }$).   We describe the evolution of the observables using the rescaled time $t/\tau$, where $\tau$ is a time scale proportional to $L$.

\subsection{Observables}
To characterize the dynamics and entanglement properties of the system, we consider several key observables derived from the fermionic correlation matrix. The observables can be easily obtained
from the correlation matrix $C_{ij} = \langle   \hat{c}^\dagger_i  \hat{c}_j  \rangle $.  For instance, 
$ \langle \hat{n}_l \rangle$ are the diagonal elements of $C$. For the trajectory-averaged bipartite entanglement entropy $\overline{S}_{L/2}$~\cite{EEntropyevolution,cauchyformula,Peschel_2009,Peschel_2004,Peschel_2003}, we first choose the left half chain as the subsystem, then get sub-correlation matrix $C_{ij}^{L/2}$ from $C$ and lastly obtain ${S}_{L/2}$ based on all eigenvalues $\{ \lambda_i \}$ of $C_{ij}^{L/2}$ with $ S_{L/2} = - {\textstyle \sum_{i} {[\lambda _i \log{(\lambda _i)}} 
+ (1-\lambda _i)\log{(1-\lambda _i)]}}$.
The trajectory-averaged velocity of particles, $\overline{v}$, as derived from the effective Hamiltonian $\hat{H}_\text{eff}$ and the correlation function $C_{ij}$, reflects the dynamics of entanglement relaxation~\cite{lzc}. The trajectory-averaged mutual information $\overline{I}_{AB}$ indicates the connection between two different subsystems, and its definition is given as $I_{AB} = S_A + S_B - S_{AB}$, with $S_A$, $S_B$ and $S_{AB}$ representing the entanglement entropies of subsystems $A$, $B$ and their union $A\cup B$, respectively. $\overline{f}_{\text{skin}}$ quantifies the time-averaged probability of the evolving state being in the ideal many-body skin state with $f_{\text{skin}} = |\left \langle \psi_{\text{skin}}  | \psi(t)  \right \rangle |^2$. The Frobenius norm $\overline{ \| D  \|}_F$ quantifies the subsystem correlations upon bipartitioning the system. Here, $D$ is extracted from the correlation matrix $C$  by selecting the submatrix $D=C_{L/2:L, 1:L/2}$, which encodes cross-partition correlations. See Appendix~\ref{sec:appendixA} for more numerical details.

\section{Numerical Results}\label{nr}
\subsection{Trajectory-averaged relaxations}\label{trajectory-averaged}

\begin{figure*}[bt]
\hspace*{-0.98\textwidth}
\centering
\includegraphics[width=0.98\textwidth]{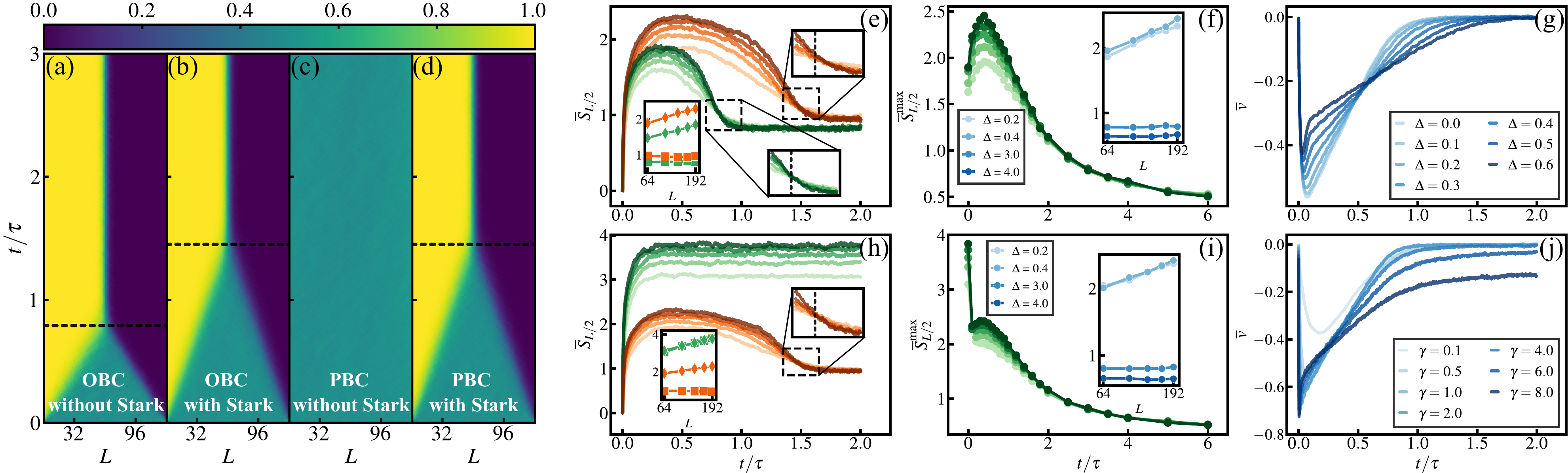} 
\caption{ Panels (a)-(d) depict the evolution of the trajectory-averaged density distribution for $L$ = $128$. (a) OBC, $\Delta$ = $0$; (b) OBC, $\Delta$ = $0.6$; (c) PBC, $\Delta$ = $0$; (d) PBC, $\Delta$ = $0.6$. Panels (e) and (h) show the evolution of the $\overline{S}_{L/2}$ for different $L$ under OBCs and PBCs. The green solid lines represent the absense of the Wannier-Stark ladder ($\Delta$ = $0$), while the orange solid lines denote its presence ($\Delta$ = $0.6$). The colors ranging from light to dark correspond to $L$ = $64$, $96$, $128$, $160$ and $192$. In the zoomed inset, we highlight the transition by the dashed black line. We also show the $\overline{S}_{L/2}$ versus $\log{L}$ at $t/\tau$ = $0.5$(diamond), $1.8$(square). Panels (f) and (i) illustrate the maximum bipartite entanglement entropy $\overline{S}^\text{max}_{L/2}$  with respect to  $L$ = $64$, $96$, $128$, $160$ and $192$ (from light to dark colors). The insets show the $\overline{S}^\text{max}_{L/2}$ versus $\log{L}$ for different $\Delta$ under OBCs and PBCs separately. Panel (g) shows the evolution of $\overline{v}$ for different $\Delta$. Panel (j) shows  the evolution of $\overline{v}$ for different $\gamma$ under $\Delta$ = $0$. Both panels (g) and (h) have a system size of $64$.
}\label{fig2}
\end{figure*}

We first investigate how the time-evolving trajectory-averaged density distribution $\overline{ \langle \hat{n}_l  \rangle }$, behaves under different boundary conditions: open boundary conditions (OBCs) and periodic boundary conditions (PBCs). Additionally, we will examine the effects of the presence and absence of a Wannier-Stark ladder. In FIG.~\ref{fig2} (a), (b), and (d), the late-time state is characterized by a notable feature: particles predominantly occupy the left half of the chain, approaching what is commonly referred to as a domain wall state, however, without a distinct boundary (hereafter termed the skin-like state). The phenomenon observed is known as the feedback-induced skin effect, as detailed in Ref.~\cite{wyp}. Specifically, under OBCs, a system under the Wannier-Stark ladder with a strength of $\Delta=0.6$ exhibits a prolonged residence in the non-equilibrium state compared to a system without the Wannier-Stark ladder ($\Delta=0$). This extended non-equilibrium duration is likely due to the restriction imposed by the Wannier-Stark ladder on particle motion.

FIG.~\ref{fig2} (c) stands out from the others in the series, as it demonstrates that  $\overline{ \langle \hat{n}_l  \rangle }$ remains relatively uniform across all lattice sites for the majority of the observation period. This uniformity can be attributed to the absence of edges and the dynamics are determined by the bulk, thus preventing the skin effect. Furthermore, when comparing FIG.~ \ref{fig2} (c) and (d), despite both under PBCs, the system in (d) does not exhibit such uniform distribution. Conversely, as time evolves the relaxation process closely mirrors the behavior observed in (b) and late-time $\overline{\langle  \hat{n}_l \rangle}$ is mainly distributed on the left of the chain. The feedback-induced skin effect still exists in this system 
despite under PBCs.
The reason is that the presence of the Wannier-Stark ladder creates a significant disparity between the first and the last on-site potentials, which generates a pseudo edge. Therefore, the particles hardly move from the first to the last site. Under PBCs, the pseudo edge plays a role similar to that of the edge under OBCs, restricting the bulk dynamics of particles.
The dashed lines shown in FIG.~\ref{fig2} (a), (c) and (d) indicate a transition point within the relaxation process, that will be further discussed in the following section.

\subsection{Entanglement relaxations and dynamical phase transitions}\label{entanglement relaxations}

Now, we delve into the dynamics of the entanglement relaxation by examining the bipartite entanglement entropy, denoted as $\overline{S}_{L/2}$, across various $L$ and  $\Delta$. Under OBCs, in FIG.~\ref{fig2} (e), $\overline{S}_{L/2}$ first sharply increases, then after the transient metastable value, slowly declines until the steady value irrespective of the presence or absence of the Wannier-Stark ladder. 
 The entanglement relaxation with the Wannier-Stark ladder ($\Delta=0.6$)
requires more time than the case without it ($\Delta=0$).
Notably, there is a crossing point during the decline of $\overline{S}_{L/2}$ for different $L$ and the same $\Delta$. We emphasize them with dashed lines in zoomed insets, called transition points. The transition point separates the log-law phase from the area-law phase. The inset of FIG.~\ref{fig2} (e)  presents several examples, illustrating the relationship between $\overline{S}_{L/2}$ and $\log{L}$ when $\Delta$ is 0.0 and 0.6, respectively. In the left of the transition point the relationship between $\overline{S}_{L/2}$ and $\log{L}$ is a linear increase; but in the right,  $\overline{S}_{L/2}$ remains constant regardless of $\log{L}$. For $\Delta=0$, the critical time (rescaled time $t_c/\tau$) is about $0.79$, which is consistent with the results in Ref.~\cite{lzc}; for $\Delta=0.6$, $t_c/\tau$ $\approx$ $1.45$. As the duration of the entanglement-relaxation process increases, the transition point is delayed,  
corresponding to the lower zone of FIG.~\ref{fig1} (b) which comprises both the log-law and area-law regions.
Our long-term observations suggest that, when $\Delta$ is sufficiently large and the evolution sufficiently slow, a  transition point may not exist. We provide examples in the Appendix \ref{big_delta} to demonstrate this viewpoint. For this case,  $\overline{S}_{L/2}$ stays in the area-law phase after an initial rapid increase. 
This novel phenomenon is demonstrated in the upper zone of FIG.~\ref{fig1} (b).

In FIG.~\ref{fig2} (h), considering $\Delta=0$ under PBCs, after a brief increase, $\overline{S}_{L/2}$  remains stable, which is log-law for the steady value. From the green solid line of FIG.~\ref{fig2} (h), $\overline{S}_{L/2}$ increases linearly with respect to $\log{L}$, and
different-time relationships are basically the same.

If $\Delta = 0.6$,  the behavior of $\overline{S}_{L/2}$ starts with a sharp increase, moves to metastable value, and finally declines to a steady value, which is the same as the case under OBCs, 
in terms of the entanglement relaxations and entanglement transitions
due to the pseudo edge.

\begin{figure*}[bt]
\hspace*{-0.99\textwidth}
\centering
\includegraphics[width=0.98\textwidth]{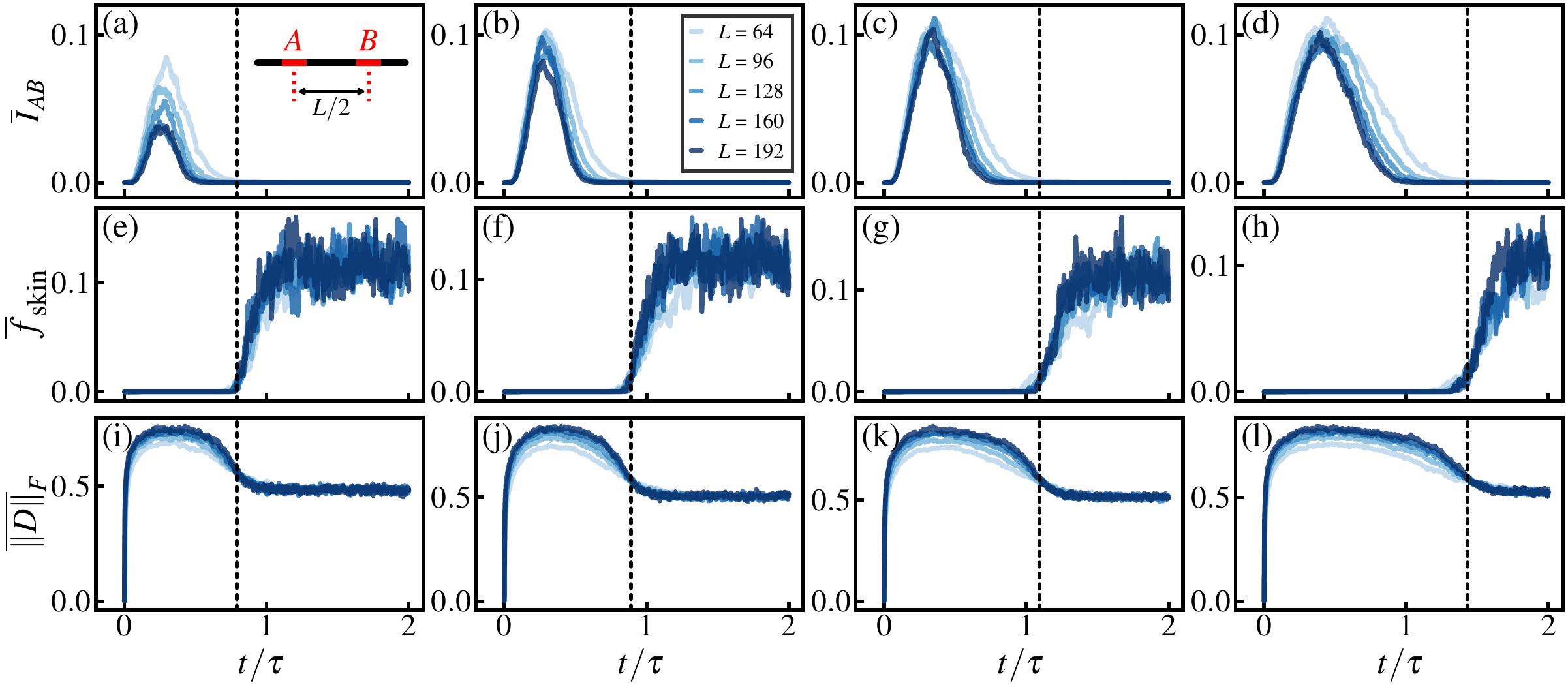} 
\caption{The evolution of different observables under various $\Delta$ values. The columns of the panels correspond to $\Delta=0.0$, $0.2$, $0.4$ and $0.6$, respectively. Panels (a)-(d) display the trajectory-averaged mutual information $\overline{I}_{AB}$. The selected subsystems $A$ and $B$, each with a size of $L/8$ and separated by $L/2$, are depicted in the schematic of panel (a). Panels (e)-(h) describe the evolution of the average probability $\overline{f}_\text{skin}$ of the evolved state being in the ideal many-body state over time. Panels (i)-(l) are about $\overline{ \| D  \|}_F $,  describing the evolution of the correlations between different subsystems under the system bipartition. The curves in the figure range from light to dark, corresponding to system sizes from small to large.
}\label{fig3}
\end{figure*}

The maximum trajectory-averaged bipartite entanglement entropy $\overline{S}_{L/2}^\text{max}$ could be found in metastable value. FIG.~\ref{fig2} (f) and (i)  illustrate how $\overline{S}_{L/2}^\text{max}$   changes with $\Delta$ for different $L$, under OBCs and PBCs respectively. In  FIG.~\ref{fig2} (f), $\overline{S}_{L/2}^\text{max}$ increases with $\Delta$ until the peak,  about $ 0.4 $, then decreases. $\overline{S}_{L/2}^\text{max}$ undergoes a transition from log-law to area-law, and we provide some illustrative instances of the different linear relationships between the
$\overline{S}_{L/2}^\text{max}$ and $\log{L}$  in the FIG.~\ref{fig2} (f). FIG.~\ref{fig2} (i) shows the PBCs case, we find that only $\Delta=0$ there is a huge difference between PBCs and OBCs because of the absence of the pseudo edge. We can also see the transition from log-law phase to area-law phase as $\Delta$ increases.

From the perspective of particle dynamics, we conduct discussions of the entanglement relaxation. Under OBCs, we have chosen the trajectory-averaged velocity of particles, represented as $\overline{v}$~\cite{lzc}, as our key observable.
In FIG.~\ref{fig2} (g), particles are initially given a velocity directed to the left. This velocity then decays linearly over time, eventually to stable zero, which marks the end of the  relaxation process. Subsequently, the particles will no longer move. The decay of $\overline{v}$ is primarily due to the edge effect. 
Furthermore, as $\Delta$ increases, the initial velocity of the particles decreases, resulting in a longer relaxation time required for the system to return to equilibrium. This indicates that $\Delta$ 
holds the bulk dynamics of the particles. The phenomenon can be explained by the fact that the smaller the initial velocity of the particles, the longer time it takes for them to reach the edge, and the edge effects exert less influence. In other words, the Wannier-Stark ladder governs the bulk dynamics and influences the relaxation time to equilibrium.

We also computed $\overline{v}$ for different values of $\gamma$.  From FIG.~\ref{fig2} (j), it is evident that under $\Delta$ = 0, when $\gamma$ is within an appropriate range (e.g.,  $\gamma$ = $0.5$, $1.0$, $2.0$ and $4.0$), the evolution of $\overline{v}$ is roughly similar. However, when $\gamma$ is particularly small or particularly large (e.g., $\gamma$ = $0.1$ or $8.0$), the evolution of $\overline{v}$ undergoes significant changes. Unlike variations in $\Delta$, altering $\gamma$ simultaneously changes both $\hat{H}_\text{eff}$ and the mean value $\langle \hat{L}_l^\dagger \hat{L}_l \rangle \gamma dt$ of the Poisson random numbers, and the effect of $\gamma$ is non-monotonic.

We investigated the time evolution of additional observables during the relaxation process, with particular emphasis on their behavior at the transition point. In FIG.~\ref{fig3}, we display three distinct observables, with each column's panels corresponding to the same $\Delta$ value. The black dashed line indicates the transition time, $t_c/\tau$, which is subsequently validated through the data collapse. 
Panels (a)–(d) illustrate the evolution of $\overline{I}_{AB}$. For various system sizes $L$, we observe that $\overline{I}_{AB}$ initially increases with time, then decreases, eventually reaching zero at $t_c/\tau$. This indicates that once the system attains a steady state, there is no correlation between subsystems $A$ and $B$.
In panels (e)–(h), we depict the evolution of $\overline{f}_\text{skin}$
, which quantifies the probability that the evolving state $| \psi(t) \rangle$ is in  the ideal many-body state  $| \psi_\text{skin} \rangle$. It is evident that before $t_c/\tau$, $\overline{f}_\text{skin}$ remains at 0, indicating that $| \psi(t) \rangle$
can never be in $| \psi_\text{skin} \rangle$. However, at $t_c/\tau$ this probability surges sharply, though it does not reach 1. This corresponds to the domain wall state without a distinct boundary,  as depicted in FIG.~\ref{fig2} (a) and (b). 
Panels (i)–(l) depict the evolution of $\overline{\|D\|}_F$. This observable, which is similar in nature to $\overline{I}_{AB}$, follows an evolution pattern akin to that of $\overline{S}_{L/2}$. As shown, a crossover occurs at $t_c/\tau$, after which the values rapidly stabilize. Taking all of the above observables into account, we find that their critical behavior is also delayed as $\Delta$ increases.

\subsection{Criticalities of the dynamical phase transitions}\label{critiality}
\begin{figure}[bt]
\hspace*{-0.48\textwidth}
\centering
\includegraphics[width=0.48\textwidth]{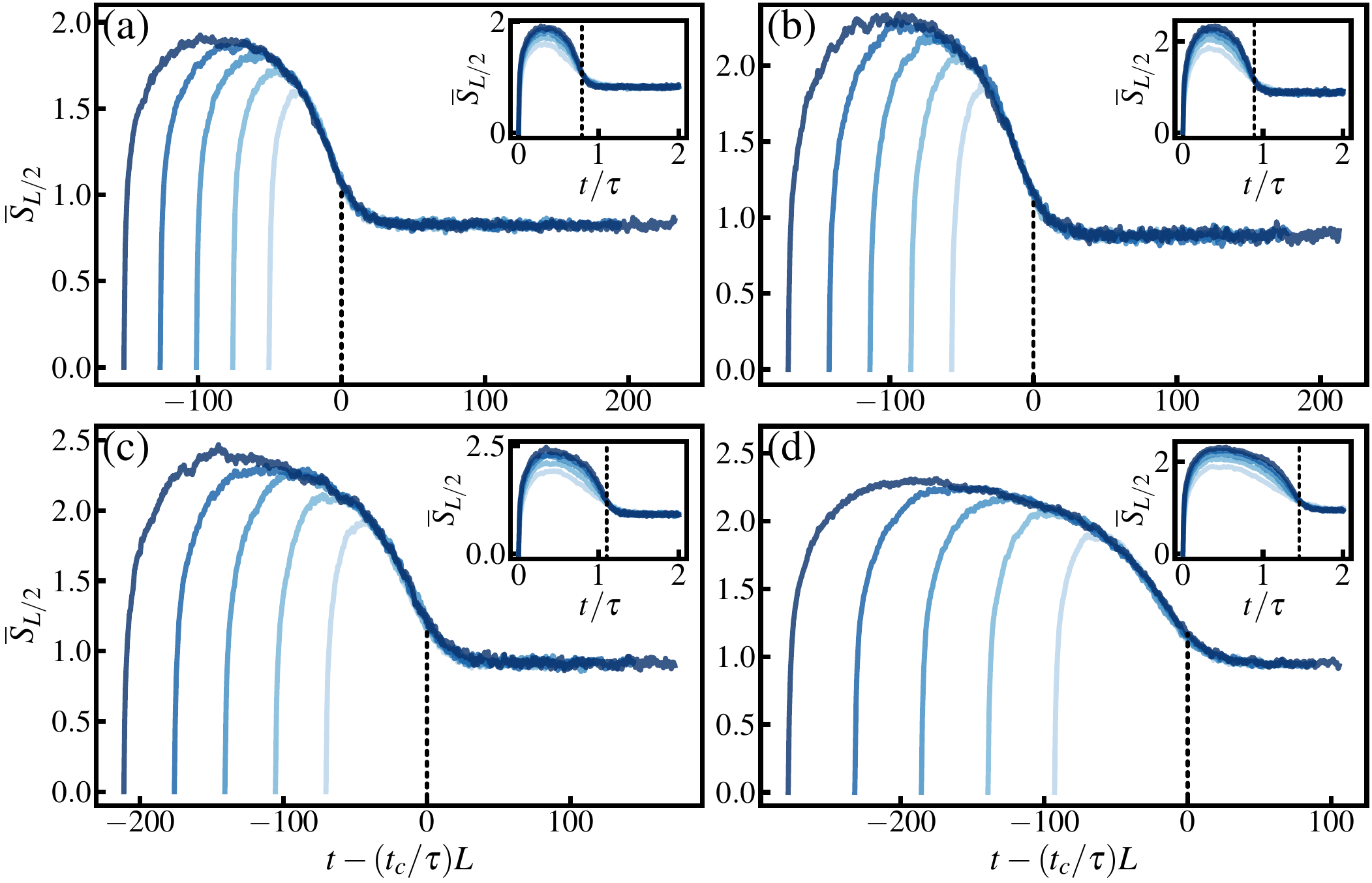} 
\caption{Data collapses for $\overline{S}_{L/2}$ with $\overline{S}_{L/2}$ versus $t-(t_c/\tau)L$. Panels (a)–(d) correspond to $\Delta$ = $0.0$, $0.2$, $0.4$ and $0.6$, while the corresponding $t_c/\tau$ values are approximately $0.79$, $0.89$, $1.10$, and $1.45$, respectively. The system sizes $L$ =  $64$, $96$, $128$, $160$ and $192$ are represented by a color gradient from light to dark, where lighter shades correspond to smaller $L$ and darker shades to larger $L$. Insets in each panel show the evolution of $\overline{S}_{L/2}$ over time for the corresponding $\Delta$. The black dashed lines in each inset represent the $t_c/\tau$ for the respective panel.}\label{fig100}
\end{figure}

For the free fermion model under open boundary conditions (OBCs), we find that the entanglement relaxation dynamics exhibits a pronounced dependence on system size. In particular, the characteristic time for entanglement growth increases approximately linearly with system size $L$, and the overall time evolution appears to follow a universal form when plotted against the rescaled time $t - (t_c/\tau)L$, as shown in Fig.~\ref{fig100}.  More importantly, as $\Delta$ increases, the critical time  $t_c/\tau$ gradually increases as well. From the panel (a) to (d),  corresponding $t_c/\tau$ values are approximately $0.79$, $0.89$, $1.10$, and $1.45$, respectively. 

This form of data collapse indicates that the entanglement dynamics is governed primarily by a system-size-dependent delay, followed by a size-independent relaxation process. Such behavior is consistent with the propagation of free fermions toward a domain-wall-like steady state, and does not exhibit any signatures of non-trivial criticality. Additionally, the presence of a Stark potential of moderate strength only prolongs the time required to reach the domain-wall-like steady state.

\section{Discussion and Conclusion}\label{argument}
Here, we have elucidated the mechanism underlying a numerically observed dynamical phase transition 
in a tilted free fermion chain subject to monitoring-feedback protocols. 
Our main findings are threefold:
\begin{enumerate}[(i)]
    \item Feedback reshapes the system's entanglement dynamics from a metastable, 
    measurement-induced log-law entangled state to an eventual area-law skin-like steady state.
    \item The resulting transition exhibits rescaling in a dimensionless time variable, $t/\tau$, where $\tau \sim L$ is the characteristic transport time in a 1D open system.
    \item A moderate Wannier-Stark ladder does not affect or alter the critical behavior of this dynamical phase transition.
\end{enumerate}

\begin{figure*}[bt]
\hspace*{-0.98\textwidth}
\centering
\includegraphics[width=0.98\textwidth]{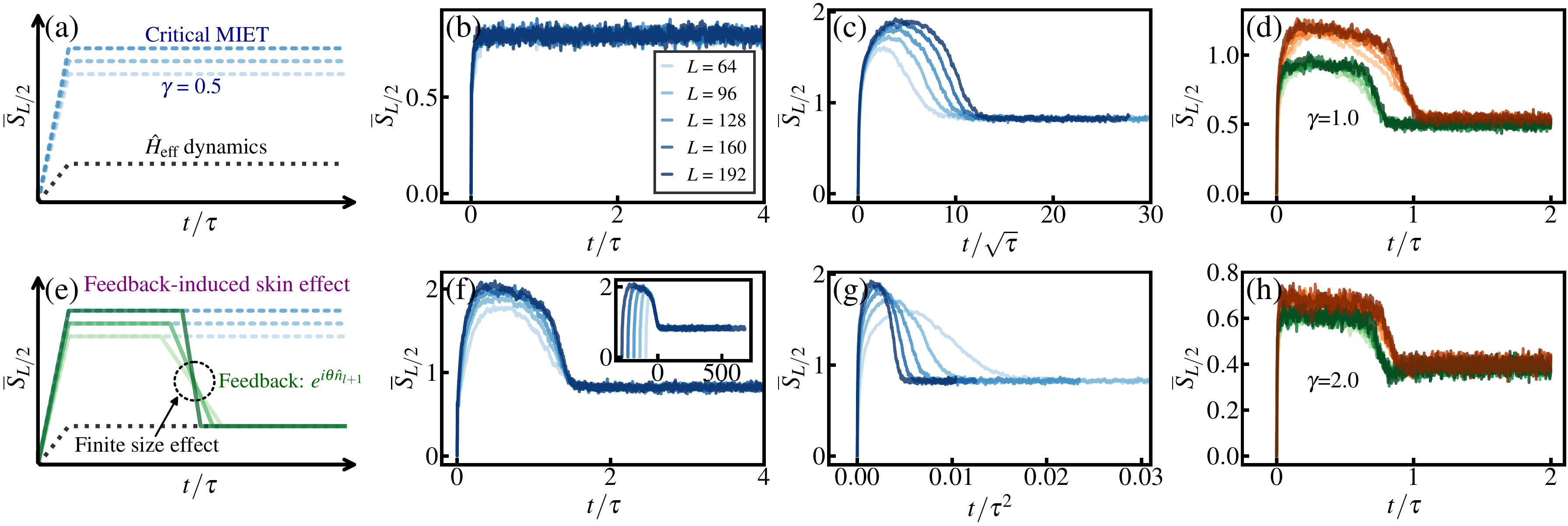} 
\caption{Panels (a) and (e) illustrate the schematic diagrams of entanglement structures with and without feedback operations, respectively. Panels (b) and (f) describe the entanglement relaxation process for another initial state. For the initial state,  the particles are mainly concentrated on the left side for panel (b) and on the right side for panel (f). The inset in (f) is the data collapse for $\overline{S}_{L/2}$ with $\overline{S}_{L/2}$ versus $t-1.46 L$. Panels (c) and (g), we rescale the time and show the evolution of $\overline{S}_{L/2}$. Panels (d) and (h) illustrate the evolution of $\overline{S}_{L/2}$ for different $\gamma$. All panels (b)-(d) and (f)-(h) are in  $\Delta=0$. 
}\label{fig5}
\end{figure*}

We begin by analyzing the limit without feedback, $\theta = 0$ in Eq.~(\ref{Eq:qj}). In this regime, the system is effectively a free-fermion setup with a MIET. At measurement rate $\gamma = 0.5$, the system resides within the critical MIET phase, where the steady-state entanglement scales logarithmically with subsystem size (log-law). Numerically, this behavior is shown by the blue dashed lines in FIG.~\ref{fig5}~(a) and (e), with deeper colors corresponding to larger system sizes. However, in the no-click limit, the dynamics are governed by the non-Hermitian effective Hamiltonian Eq.~(\ref{Eq:Heff}), which drives the system into a skin-like steady state exhibiting area-law entanglement. This purely non-Hermitian behavior is illustrated by the black dashed lines in FIG.~\ref{fig5}~(a) and (e).

Separately, introducing the feedback operator Eq.~(\ref{Eq:qj}) with $\theta=\pi$ balances these two extremes. The feedback acts to erase the noise of quantum jumps, thereby pushing the system toward the area-law entangled, non-Hermitian skin-like steady state. However, the feedback effect is not instantaneous. Initially, the system remains in a metastable log-law entangled regime; only after a characteristic relaxation time does the feedback reshape the evolving state into the final skin-like state. This two-stage process is shown schematically by the solid green lines in FIG.~\ref{fig5}~(e) and numerically in FIG.~\ref{fig2}~(e) (green lines). Early on, the system exhibits log-law entanglement, then crosses over to the area-law entangled steady state. The finite-size crossing in the time domain thus emerges naturally as a feedback-induced dynamical phase transition. Additionally, the lifetime of the metastable log-law regime is determined by how quickly the feedback pumping reshapes the system into the skin-like steady state. If the initial state is already similar to the final steady state, relaxation occurs swiftly, and the metastable window is narrow. Conversely, if the initial state is orthogonal, a longer metastable regime is observed. FIG.~\ref{fig5}~(b) illustrates the entanglement dynamics starting from the effective Hamiltonian's skin-like steady state: no metastable regime appears, as the system is already at equilibrium under the non-Hermitian dynamics. In contrast, FIG.~\ref{fig5}~(f) shows the opposite case, where the initial state is a reverse skin state, fully orthogonal to the final steady state. The metastable window is significantly prolonged, and the dynamical transition time increases to $t_c/\tau \approx 1.46$, compared to a shorter time in the N\'eel-state case [FIG.~\ref{fig2}~(e), green lines]. 

To understand the critical behavior, we note that $t/L$ (proposed in Refs.~\cite{wyp,fx,lzc,PhysRevB.110.144305}) are dimensionally confusing, as it does not carry units of time. Instead, we identify $t/\tau$ as a more natural scaling variable, where $\tau$ is the time for particles to traverse a 1D open chain of length $L$. As $\tau \sim L$, the ratio $t/\tau$ is dimensionless and suitable for the data collapse. When other nonlinear time rescalings are attempted [FIG.~\ref{fig5}~(c) and (g)], signatures of the dynamical phase transition disappear, confirming that $t/\tau$ is the appropriate scaling parameter. Moreover, at higher measurement rates $\gamma$ [FIG.~\ref{fig5}~(d) and (h)], the metastable state itself becomes area-law, rather than log-law, but we anticipate the same feedback mechanism to induce a crossover between two distinct states in the entanglement evolution. Observing this transition in practice, however, is more challenging because stronger quantum-jump noise obscures the crossover. 

It worth to notice that the Wannier-Stark ladder preserves the critical behavior of this dynamical phase transition. As discussed in Ref.~\cite{lzc}, the localization of the effective Hamiltonian only monotonically influences the time it takes for particles to be transported to the boundary. The stronger the localization, the longer the transport time. This suggests that the criticality of this dynamical phase transition is not entirely determined by the dynamics of the effective Hamiltonian.

To sum up, we systematically investigate a dynamical phase transition in monitored free fermion chain subject to feedback control. We study the evolution of the particle density distribution and entanglement entropy under different boundary conditions (open and periodic) and analyze the impact of the Wannier-Stark ladder. Firstly, we identify and characterize the feedback-induced skin effect, where the trajectory-averaged particle density becomes localized on one side of the lattice at late times. Introducing a Wannier-Stark ladder significantly prolongs the residence time in this non-equilibrium state, as the potential gradient restricts particle transport. Under periodic boundary conditions, this gradient creates a pseudo-edge, leading to a similar localization effect even without physical edges. Secondly, by analyzing the trajectory-averaged entanglement entropy, we observe a clear two-stage dynamical process: initially, entanglement rapidly grows into a metastable regime characterized by logarithmic scaling (log-law), and subsequently decays slowly into a final steady state exhibiting area-law entanglement. This transition between log-law and area-law behaviors occurs at a distinct dynamical transition point, confirmed by examining multiple observables, including subsystem mutual information, probability of the evolving state being in the ideal skin state,  and subsystem correlation norms.  Importantly, we provide an intuitive and graphical explanation for the mechanism driving this dynamical phase transition: feedback gradually suppresses quantum jump noise, reshaping the system’s steady state from a measurement-induced metastable log-law entangled state into a non-Hermitian skin-like area-law state. Moreover, we discuss how the presence of the Wannier-Stark ladder affects the dynamics: while it significantly delays the relaxation process, it does not alter the critical behavior of the observed dynamical phase transition. Finally, we explore the fate of this dynamical transition in the presence of a strong Wannier-Stark ladder. We propose that if the gradient is sufficiently large, the system may directly relax into an area-law regime without undergoing a clear dynamical transition.

This work further complements the dynamical phase transition induced by the feedback-induced skin effect previously proposed in Ref.~\cite{lzc}. Additionally, experimentally observing this dynamical phase transition requires a platform capable of precise real-time feedback control, single-particle resolution measurements, and tunable potential gradients, which are likely achievable in ultracold atom optical lattices~\cite{Georgescu_2014} or ion-trap systems~\cite{Blatt2012,Georgescu_2014,fossfeig2024progresstrappedionquantumsimulation}. However, implementing high-precision, rapid feedback loops for dynamically adjusting system parameters, as well as accurately measuring and controlling entanglement entropy and particle density distributions, remains challenging.


\begin{acknowledgements}
We thank Ze-Chuan Liu, Yu-Peng Wang, and Shan-Zhong Li for discussions. H.-Z. Li appreciates the guidance of Ching Hua Lee and Xue-Jia Yu. J.-X. Zhong acknowledges the National Natural Science Foundation of China (Grant No.~12374046 and No.~11874316), the National Basic Research Program of China (Grant No. 2015CB921103), and the Program for Changjiang Scholars and Innovative Research Teams in Universities (Grant No. IRT13093).
 
\end{acknowledgements}


\appendix

\section{Numerical Method}
\label{sec:appendixA}
\subsection{Numerical simulation for evolution of the state}
We  currently employ Eq.(\ref{Eq:SSE}) to evolve  $\left | \psi(t)  \right \rangle$  treated as a Slater determinant state. We can depict $\left | \psi(t)  \right \rangle$ with a matrix $U\left(t\right)$ of size $L \times N$ at time $t$, 

\begin{equation} \label{Eq:MatrixForState}
\left | \psi(t)  \right \rangle = 
\prod_{n=1}^{N}{\left( \sum_{l=1}^{L}{[U(t)]_{ln} } \hat{c}_l^\dagger\right) } \left |0  \right \rangle,
\end{equation}
where $L$ is the system size and $N$ is the number of particles, which satisfy $N = L/2$. $\left |0  \right \rangle$ stands for vacuum state. Moreover, any column $U_n$ of $U\left(t\right)$ can represent the $n$-th single-particle state,
\begin{equation} \label{Eq:ColumnForState}
\left | \psi_n  \right \rangle = 
\sum_{l=1}^{L}{\left( [U_n]_{l}  \hat{c}_l^\dagger\right) \left |0  \right \rangle}.
\end{equation}

After getting the matrix form $H_\text{eff}$ of effective Hamiltonian $\hat{H}_\text{eff}$, non-Hermitian evolution can be written as
\begin{equation} \label{Eq:NonHermitianEvolution}
U\left( t+\delta t\right) = e^{-iH_\text{eff}\delta t}U\left( t \right).
\end{equation}
Following the evolution, we require the application of QR decomposition to ensure the orthogonality among different single-particle states and the normalization of each individual single-particle state.

Now we consider the continuous measurement with feedback and numerical simulation of the quantum jump, i.e.,
\begin{equation}
\left | \psi\left(t\right)  \right \rangle \longrightarrow \frac{\hat{L}_l \left | \psi\left(t\right)  \right \rangle }{\sqrt{\langle \hat{L}_l^\dagger  \hat{L}_l \rangle } }.
\end{equation}
The quantum jump probability can be written as
\begin{equation}
p_l = \langle \hat{L}_l^\dagger \hat{L}_l   \rangle  \gamma \delta t.
\end{equation}
Using Wick’s theorem,
\begin{equation}
\langle \hat{d}_l^\dagger \hat{d}_l \hat{d}_l^\dagger \hat{d}_l  \rangle = 
\langle \hat{d}_l^\dagger \hat{d}_l  \rangle  \langle \hat{d}_l^\dagger \hat{d}_l  \rangle + 
\langle \hat{d}_l^\dagger \hat{d}_l  \rangle  \langle \hat{d}_l\hat{d}_l^\dagger \rangle,
\end{equation}
and the fermionic anticommutation relations,
\begin{equation}
\{  \hat{c}_i, \hat{c}_j^\dagger \} = \delta _{ij},
\end{equation}
we get the probability of quantum jump,
\begin{equation}
p_l = \langle  \hat{d}_l^\dagger  \hat{d}_l  \rangle \gamma \delta t =
\left\langle\psi(t)\right|  \hat{d}_l^\dagger  \hat{d}_l \left | \psi(t)  \right \rangle \gamma  \delta t =
\| \hat{d}_l \left |\psi(t)  \right \rangle \|^2 \gamma \delta t.
\end{equation}

\begin{figure}[bt]
\hspace*{-0.5\textwidth}
\centering
\includegraphics[width=0.5\textwidth]{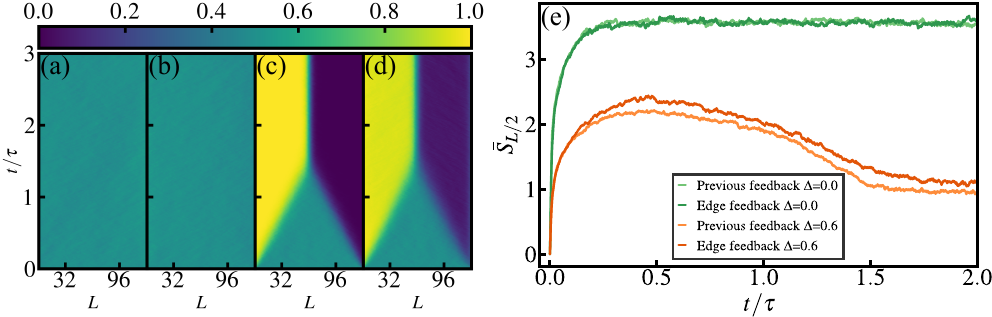} 
\caption{PBCs, the trajectory-averaged density distribution evolution with respect to rescaled time for $L=128$ . (a) previous feedback, $\Delta=0$; (b) edge feedback, $\Delta=0$; (c) previous feedback, $\Delta=0.6$; (d) edge feedback, $\Delta=0.6$. In panel (e), $\overline{S}_{L/2}$ evolves with respect to rescaled time for different feedback and $\Delta$.}\label{fig6}
\end{figure}

Note that the linear expansion of $\hat{d}_l$ can be written as
\begin{equation}
\hat{d}_l =  \sum_{l}^{L} {[\textbf{d}]_l \hat{c}_l},
\end{equation}
where $\textbf{d}$ is a row vector derived from the relation
\begin{equation}
\hat{d}_l = \frac{\hat{c}_l + i\hat{c}_{l+1}}{\sqrt{2}}.
\end{equation}
To simulate the quantum jump, we can find a single-particle state which satisfies $ \textbf{d}  U_n \ne 0$ and $n > 1$, then move $U_n$ to the first due to gauge freedom and transform the rest columns of $U(t)$ accordingly,
\begin{equation} \label{Eq:Schmidt}
U_n' = U_n - \frac{\textbf{d} U_n}{\textbf{d} U_1} U_1 .
\end{equation}
The result is 
\begin{equation} \label{Eq:ww}
U'\left( t \right) = 
\begin{bmatrix}
    {\textbf{d}}^\dagger & U_2' & U_3' & \cdots & U_N'
\end{bmatrix}.
\end{equation}
Finally, we use the QR decomposition for $e^{i\pi\mathcal{N}_{l+1}  }U'(t)$, and $\mathcal{N}_{l+1}$ is the matrix form of particle number operator.

\subsection{trajectory-averaged observables}
 The Gaussian state allows for the extraction of observables from the correlation matrix $C_{ij} =  \langle \hat{c}_i^\dagger  \hat{c}_j  \rangle =\left [ U(t)U^\dagger (t) \right ]^T_{ij}$. Therefore, density distribution $\left \langle \hat{n}_l \right \rangle$ can be obtained through the diagonal elements of the correlation matrix.

When calculating the bipartite entanglement entropy $S_{L/2}$, we choose the sub system  $[1, L/2]$ and then obtain all the eigenvalues \{$\lambda_i$\} of sub-correlation matrix $C_{ij}^{L/2}$. $S_{L/2}$  is given by ~\cite{EEntropyevolution,cauchyformula,Peschel_2009,Peschel_2004,Peschel_2003}
\begin{equation} \label{Eq:SL2}
S_{L/2} = -\sum_{i=1}^{L/2}{\lambda_i\log{(\lambda_i)} + (1-\lambda _i)\log{(1-\lambda _i)}} .
\end{equation}

When calculating the velocity of particles under OBCs, taking into account  that the nearest-neighbor distance is normalized, we can get the average position for particles  depicted by $ \langle \hat{x}  \rangle =  {\textstyle \sum_{l}^{L} l \langle \hat{n}_l   \rangle } /N$. Furthermore, the average velocity per particle $v$ equals $d \langle \hat{x}  \rangle/dt$ . The master equation, 
\begin{equation} \label{Eq:master}
\frac{d\rho _t}{dt} =-i\hat{H}_\text{eff} \rho _t + i  \rho _t \hat{H}_\text{eff}^\dagger + \gamma \sum_{m}^{L-1}
{\hat{L}_m \rho _t \hat{L}_m^\dagger  } ,
\end{equation}
which governs the evolution of the density matrix $\rho _t$. Using $ \langle \hat{n}_l   \rangle$ to replace $\rho _t$, we have 
\begin{equation} 
 \frac{d\left \langle \hat{n}_l \right \rangle }{dt} 
=-i\left \langle\hat{n}_l \hat{H}_\text{eff} \right \rangle+ 
i\left \langle  \hat{H}_\text{eff}^\dagger \hat{n}_l\right \rangle + 
\gamma \sum_{m}^{L-1}{\left \langle  \hat{L}_m^\dagger  \hat{n}_l \hat{L}_m  \right \rangle}.
\end{equation}

\begin{figure}[bt]
\hspace*{-0.48\textwidth}
\centering
\includegraphics[width=0.48\textwidth]{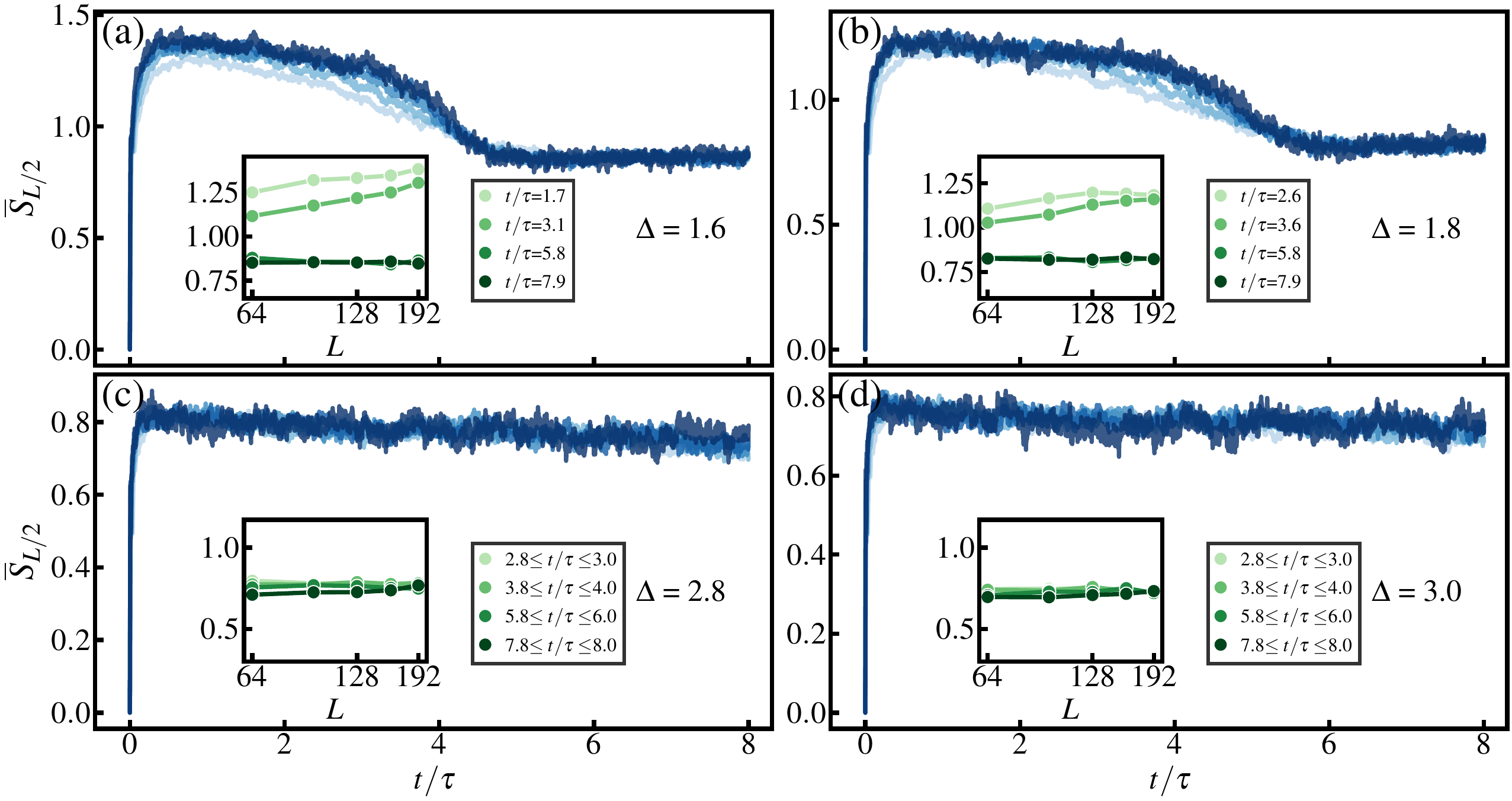} 
\caption{ $\overline{S}_{L/2}$ versus rescaled time $t/\tau$. (a) $\Delta$ = $1.6$; (b) $\Delta$ = $1.8$; (c) $\Delta$ = $2.8$; (d) $\Delta$ = $3.0$. The insets show the $\overline{S}_{L/2}$ versus $\log{L}$, either at different time points or averaged over time intervals.}\label{fig7}
\end{figure}

Finally, we employ Wick theorem and get the average velocity per particle,

\begin{equation} 
\begin{aligned}
v &\ = 
\frac{1}{N}\sum_{l}{i \left( l\left \langle \hat{H}_\text{eff}^\dagger \hat{n}_l \right \rangle - l\left \langle \hat{n}_l \hat{H}_\text{eff}\right \rangle \right) }\\
&\ + \frac{1}{N} \sum_{l, m}{l\left[ 
\frac{1}{2}\left(\delta_{lm} + \delta_{l, m+1} \right) \gamma \left \langle \hat{d}_m^\dagger \hat{d}_m \right \rangle \right]}\\
&\ -  \frac{1}{N} \sum_{l, m}{il  \left \langle \hat{n}_l \right \rangle\left \langle \hat{H}_\text{eff}^\dagger -\hat{H}_\text{eff}\right \rangle}\\
&\ + \frac{1}{N} \sum_{l, m}{l\gamma  \left \langle \hat{d}_m^\dagger \hat{c}_l \right \rangle
\left \langle \hat{c}_l^\dagger \hat{d}_m \right \rangle.
}
\end{aligned}
\end{equation}

When calculating the mutual information $I_{AB}$, we first define two subsystems, each of size $L/8$, separated by a distance of $L/2$, and label them $A$ and $B$. Using the same method previously employed for computing $\overline{S}_{L/2}$, we then determine their individual entanglement entropies, $S_A$ and $S_B$, as well as the entanglement entropy of the combined subsystem, $S_{AB}$. The mutual information is given by $I_{AB} = S_A+S_B-S_{AB}$.

When calculating the $f_\text{skin}$, we use the matrix $U(t)$
, which describes $| \psi(t) \rangle$, and the matrix $U_\text{skin}$
, which describes $| \psi_\text{skin} \rangle$, to perform the following calculation:
$|U_\text{skin}^\dagger U(t)|^2$.

When computing $\|D\|_F$, we select the submatrix $D=C$[$L/2$: $L$, $1$: $L/2$] from the correlation matrix , and then calculate the Frobenius norm of $D$.

\begin{figure}[bt]
\hspace*{-0.48\textwidth}
\centering
\includegraphics[width=0.48\textwidth]{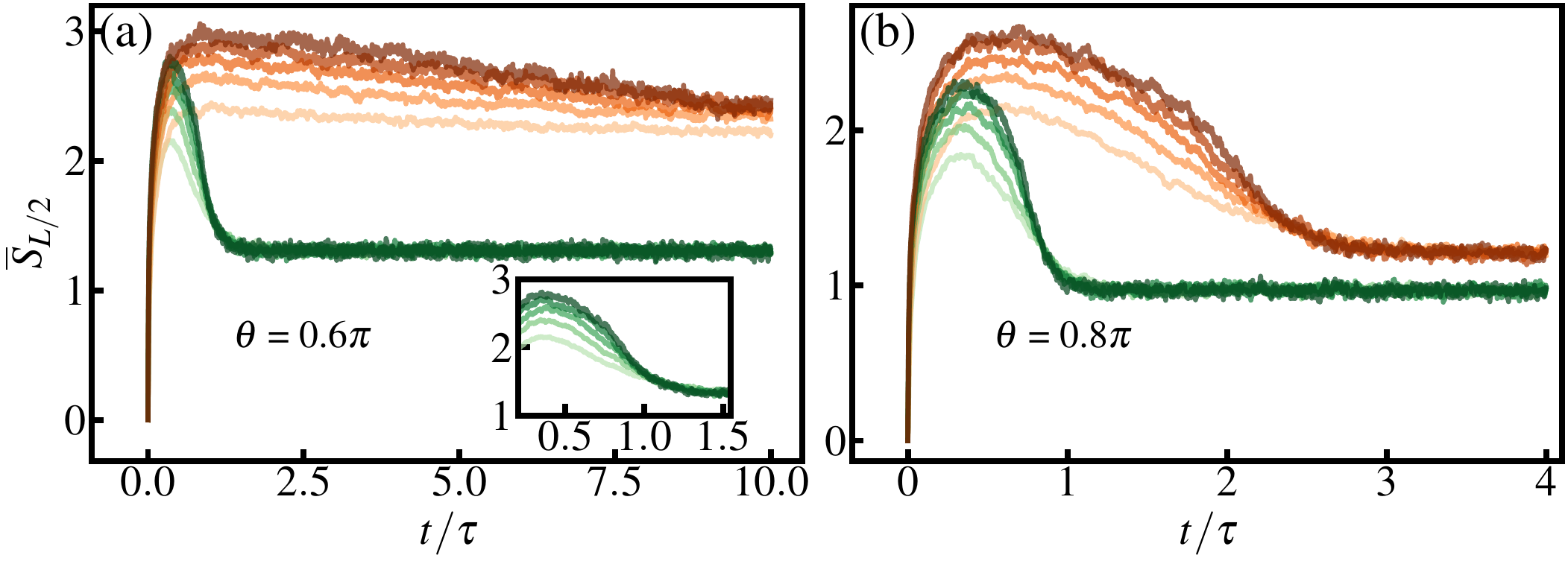} 
\caption{ $\overline{S}_{L/2}$ evolves under different $\theta$. The green solid line represents 
$\Delta$ = $0$, and the orange solid line represents 
$\Delta$ = $0.6$. The color gradient from light to dark corresponds to increasing $L$ values from smaller to larger.  (a) $\theta=0.6\pi$; (b) $\theta=0.8\pi$. The inset in panel (a) provides an enlarged view of the region surrounding the transition point at $\Delta = 0$.}\label{fig8}
\end{figure}

\section{Feedback protocol with boundary effect for the Wannier-Stark ladder}\label{edgefeedback}

In this section, we focus on a different feedback protocol, considering the boundary effects with PBCs. This is because a chain with Wannner-Stark ladder exhibits peculiar phenomena under PBCs, behaving similarly to OBCs and showing anomalous bulk dynamics. We incorporate the edge, considering the edge feedback protocol, and the quantum jump operator is written as,

\begin{equation}
\hat{L}_l' = 
\begin{cases}
  e^{i\theta\hat{n}_{l+1}}\hat{d}_l^\dagger \hat{d}_l & \text{ if } l = 1,2,...,L-1,\\
   e^{i\theta\hat{n}_{1}}\hat{d}_L^\dagger \hat{d}_L & \text{ if } l=L,
\end{cases}
\end{equation}
where $\hat{d}_L^\dagger =  (  \hat{c}_L^\dagger - i\hat{c}_1^\dagger )/\sqrt{2}$. The edge feedback protocol (hereafter referred to as ``edge feedback") differs from the previous feedback protocol, which will be abbreviated as ``previous feedback" in this section. The quantum jump with the edge feedback accounts for the additional ``left moving" from the initial site to the final one.

From FIG.~\ref{fig6} (a) and (b), $\overline{ \left \langle \hat{n}_l  \right \rangle }$ for previous and edge feedback are the same without the Wannier-Stark ladder. The bipartite entanglement entropy $\overline{S}_{L/2}$ are the same for different feedback likewise, from green solid lines in FIG.~\ref{fig6} (e). Comparing FIG.~\ref{fig6} (c) and (d), despite some similarities, the steady state in (d) is not completely occupied in the left part, and the right part shows a sparse density distribution. It is evident from the orange solid lines in FIG.~\ref{fig6} (e) that the $\overline{S}_{L/2}$ for the edge feedback is consistently higher than the other. The reason for the phenomenon is that edge feedback affects the pseudo-edge effect, which is caused by the Wannier-Stark ladder. A minority of particles can cross the pseudo edge from the first to the last site with the aid of the quantum jump with the edge feedback. If $\Delta=0$, the absence of the pseudo edge occurs, so observables remain the same for different feedback. 

\section{The edge of dynamical phases in phase diagram}\label{big_delta}
In this section, we discuss the ambiguous boundary between the log-law phase and the upper area-law phase ($\Delta$ = $1.6$, $1.8$), as well as the top edge of the phase diagram ($\Delta$ = $2.8$, $3.0$). In both FIG.~\ref{fig7} (a) and (b), the transition from the log-law phase to the area-law phase occurs at significantly later times compared to the cases of $\Delta$ = $0.0$ and $0.6$, seeing FIG.~\ref{fig2} (e). The insets detail the behavior before and after the crossing point, clearly illustrating the transition. 
In FIG.~\ref{fig7} (c) and (d), no transition is observed within the extended time range we selected. The insets present data averaged over various time intervals, indicating that $\overline{S}_{L/2}$ essentially remains unchanged as $\log{L}$ increases.

Subsequently, we employ the fitting function
$\overline{S}_{L/2} = a \log{L} + b$ to analyze the $\overline{S}_{L/2}$ versus $\log{L}$ data for various $\Delta$ values with the aim of determining the coefficient $a$. For $\Delta = 0.0$, $0.2$, $0.4$, $0.6$, $1.6$, and $1.8$, when $t/\tau$ is set to $0.5$, $0.5$, $0.5$, $0.5$, $3.1$, and $3.6$, respectively, we obtain $a \approx 0.33$, $0.43$, $0.42$, $0.37$, $0.16$, and $0.13$.  However, when applying the same fitting procedure to the data for $\Delta=2.8$ and $3.0$, the resulting $a$ values are all less than $10^{-2}$, and the plot of $\overline{S}_{L/2}$ versus $\log{L}$ shows only minimal variation. We note that this behavior is similar to that observed for $\Delta=0.0$ at $t/\tau=1.8$, where the system is in the area-law phase. Therefore, combined with the previously noted observation that no transition was observed over an extended time period, we suggest that for large $\Delta$, the system remains in the area-law phase.

\section{The effect of variations in $\theta$ on entanglement relaxation.}\label{theta}

In this section, we attempt to use feedback with different 
$\theta$ values under OBCs to investigate the effect of 
$\theta$ variations on the evolution of $\overline{S}_{L/2}$. In FIG.~\ref{fig8} (a), with 
$\theta=0.6\pi$, when $\Delta=0$
 the corresponding 
$t_c/\tau$ $\approx$ $0.98$; however, when 
$\Delta = 0.6$, based on the continuously decreasing 
$\overline{S}_{L/2}$ over time, it can be inferred that its transition point will occur after $t_c/\tau$ $=$ $10$.
 In FIG.~\ref{fig8} (b), with $\theta=0.8$, when $\Delta=0$, $t_c/\tau \approx 0.83$; however, when 
$\Delta=0.6$, the time corresponding to $t_c/\tau$ is significantly larger than that of the orange lines in FIG.~\ref{fig2} (e). This indicates that the system in the Wannier-Stark ladder  is more sensitive to variations in $\theta$.

\bibliography{Refs}
\end{document}